\documentclass[aps, prl, showpacs, twocolumn,superscriptaddress]{revtex4-1}

\usepackage{dcolumn}
\usepackage{graphicx}
\usepackage{bm, amsmath, amssymb, textcomp}
\usepackage{hyperref}

\begin{document}

\title{Phonon-assisted optical absorption in silicon from first principles}

\author{Jesse Noffsinger}
\affiliation{Department of Physics, University of California, Berkeley, California 94720, USA}
\affiliation{Materials Sciences Division, Lawrence Berkeley National Laboratory, Berkeley, California 94720, USA}
\author{Emmanouil Kioupakis}
\affiliation{Materials Department, University of California, Santa Barbara, California 93106, USA}
\affiliation{Department of Materials Science and Engineering, University of Michigan, Ann Arbor, Michigan 48109, USA}
\author{Chris G. Van de Walle}
\affiliation{Materials Department, University of California, Santa Barbara, California 93106, USA}
\author{Steven G. Louie}
\affiliation{Department of Physics, University of California, Berkeley, California 94720, USA}
\affiliation{Materials Sciences Division, Lawrence Berkeley National Laboratory, Berkeley, California 94720, USA}
\author{Marvin L. Cohen}
\affiliation{Department of Physics, University of California, Berkeley, California 94720, USA}
\affiliation{Materials Sciences Division, Lawrence Berkeley National Laboratory, Berkeley, California 94720, USA}

\date{\today}

\begin{abstract}
The phonon-assisted interband optical absorption spectrum of silicon is calculated at the 
quasiparticle level entirely from first principles. We make use of the Wannier 
interpolation formalism to determine the quasiparticle energies, as well as the 
optical transition and electron-phonon coupling matrix elements, on fine grids 
in the Brillouin zone. The calculated spectrum near the onset of indirect 
absorption is in very good agreement with experimental measurements for a 
range of temperatures.  Moreover, our method can accurately determine the 
optical absorption spectrum of silicon in the visible range, an important 
process for optoelectronic and photovoltaic applications that cannot be 
addressed with simple models. The computational formalism is quite general 
and can be used to understand the phonon-assisted absorption processes in 
general.


\end{abstract}

\maketitle

\begin{figure}
\includegraphics[scale=1.2]{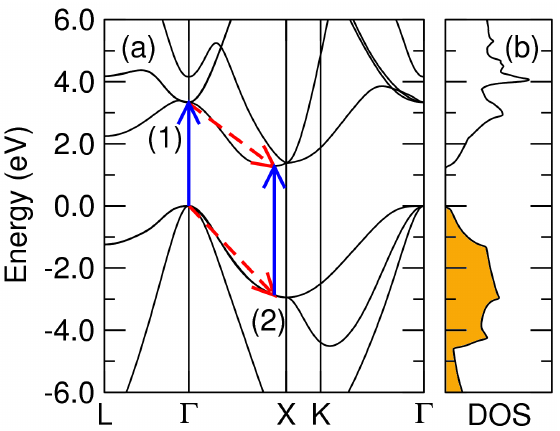}
\caption{
\label{fig:fig1}
(a) Quasiparticle band structure of silicon calculated within the GW approximation
and interpolated with the Wannier formalism. The arrows indicate the 
lowest-energy phonon-assisted optical absorption processes across the 
indirect band gap. Solid lines denote optical transitions, while dashed 
lines correspond to electron-phonon scattering events. The two terms, 
$\bm{S}_1$ and $\bm{S}_2$, of Eq.~\ref{eq:alpha2} are represented by 
paths (1) and (2) respectively. (b) Density of electronic states versus 
the quasiparticle energy. The density of states of the occupied bands has been highlighted.
}
\end{figure}

The phonon-assisted absorption of light in materials is an 
important optical process both from a fundamental and from 
a technological point of view. Intraband light absorption by free 
carriers in metals and doped semiconductors requires the 
additional momentum provided by the lattice vibrations 
or defects,
while phonon-assisted processes determine the onset of 
absorption in indirect-band-gap semiconductors (Fig.~\ref{fig:fig1}).  
Moreover, the value of the direct band gap in silicon 
(3.4 eV\cite{PhysRevB.1.2668}) is large and precludes 
optical absorption in the visible. However, silicon is a 
commercially successful photovoltaic material because of the
indirect optical transitions that enable photon capture in 
the spectral region between the indirect (1.1 eV\cite{Welber19751021}) 
and direct band gaps.

Despite their importance, at present, only a very limited number 
of first-principles studies of phonon-assisted optical 
absorption spectra exist.
Ab initio calculations of direct optical absorption spectra including excitonic effects
have already been performed for Si and other bulk semiconductors \cite{PhysRevB.57.R9385, *PhysRevLett.80.4510, *PhysRevLett.81.2312}
and the underlying methodology is presently well established \cite{PhysRevB.62.4927,RevModPhys.74.601}.
Phonon-assisted absorption studies are more involved, however, and the associated computational
cost is much higher than the direct case.
The calculation of the indirect absorption 
coefficient involves a double sum over $\bm{k}$-points 
in the first Brillouin zone (BZ)
to account for all initial and final electron states.
In addition, these sums must be performed with a very fine sampling 
of the zone to get an adequate spectral resolution. The 
computational cost associated with these BZ sums is in fact 
prohibitive with the usual methods. Phonon-assisted absorption 
calculations have been done for the special case of free-carrier 
absorption in semiconductors\cite{PhysRevB.81.241201}, where the carriers are initially 
limited to a region near the $\Gamma$ point of the first BZ, but a 
full calculation using brute-force methods for the general case 
remains beyond the reach of modern computers.

The difficulty of zone-integral convergence can be addressed with the 
maximally-localized Wannier function interpolation 
method\cite{MarzariVanderbilt97,*SouzaMarzariVanderbilt01,*wannier90}.
Using this technique, the quasiparticle energies\cite{PhysRevB.82.245203}
and optical transition matrix elements\cite{PhysRevB.75.195121,*PhysRevB.74.195118}
can be interpolated to arbitrary points in the BZ at a minimal computational 
cost. Moreover, this interpolation method has been 
generalized\cite{PhysRevB.76.165108,*PhysRevLett.99.086804}
to obtain the electron-phonon coupling matrix elements between
arbitrary pairs of points in the first BZ.

In this Letter, we developed a first-principles computational method, based 
on a Wannier-Fourier interpolation formalism, to calculate the phonon-assisted 
optical absorption spectra of materials from first principles and applied it to the case of 
interband absorption in silicon.
The calculated 
spectra near the absorption onset are in very good agreement with experimental 
results for a range of temperatures. Moreover, we were able to reproduce the 
absorption spectrum in the energy range between the indirect and 
direct band gaps (1.1 -- 3.4 eV), a spectral region that cannot be accessed 
by standard model calculations. This region covers the entire visible spectrum and is 
important for optoelectronic applications.
The computational formalism is quite general and can be used to predict and analyze the
phonon-assisted optical absorption spectrum of any material.

To calculate the phonon-assisted absorption coefficient,
we use the Fermi's golden rule expression\cite{BassaniParravicini,PhysRevB.81.241201}:
\begin{align}{\label{eq:alpha2}}
\alpha(\omega) = & 2\frac{4 \pi^2 e^2}{\omega c n_r(\omega)}\frac{1}{V_{\text{cell}}}\frac{1}{N_{\bm{k}} N_{\bm{q}}}
\sum_{\nu i j \bm{k} \bm{q}}\left| \bm{\lambda} \cdot \left( \bm{S}_1+\bm{S}_2\right) \right| ^2 \nonumber \\
 & \times P  \delta(\epsilon_{j,\bm{k}+\bm{q}}-\epsilon_{i\bm{k}}-\hbar\omega \pm \hbar\omega_{\nu \bm{q}} ),
\end{align}
where $\hbar\omega$ and $\bm{\lambda}$ are the energy and polarization of the photon 
and $n_r(\omega)$ is the refractive index of the material at frequency $\omega$.
The generalized optical matrix elements, $\bm{S}_1$ and $\bm{S}_2$, are 
given by
\begin{align}
\label{eq:S1} \bm{S}_1(\bm{k},\bm{q}) = & \sum_m \frac{  \bm{v}_{im}(\bm{k}) g_{mj,\nu}(\bm{k},\bm{q}) }{\epsilon_{m\bm{k}}-\epsilon_{i\bm{k}}-\hbar\omega + i\Gamma_{m,\bm{k}}}, \\
\label{eq:S2} \bm{S}_2(\bm{k},\bm{q}) = & \sum_m\frac{ g_{im,\nu}(\bm{k},\bm{q}) \bm{v}_{mj}(\bm{k}+\bm{q}) }{\epsilon_{m,\bm{k}+\bm{q}}-\epsilon_{i\bm{k}}\pm \hbar\omega_{\nu \bm{q}} + i\Gamma_{m,\bm{k}+\bm{q}}},
\end{align}
and correspond to the two possible paths of the indirect absorption process (Fig.~\ref{fig:fig1}).
They are determined in terms of the velocity ($\bm{v}$) and electron-phonon coupling ($g$) matrix elements,
as well as  the real ($\epsilon_{n\bm{k}}$) and imaginary 
($\Gamma_{n\bm{k}}$) parts of the quasiparticle self-energies.
The factor $P$ accounts for the carrier and phonon statistics,
\begin{align}
P = \left(n_{\nu \bm{q}} +\frac{1}{2} \pm \frac{1}{2} \right)(f_{i\bm{k}}-f_{j,\bm{k}+\bm{q}}). \nonumber
\end{align}
The upper (lower) sign corresponds to phonon emission (absorption).

The Kohn-Sham eigenvalues were calculated within the local density 
approximation (LDA)\cite{CeperleyAlder80,*PerdewZunger81} to density 
functional theory using a plane-wave pseudopotential 
approach\cite{IhmZungerCohen79} with a kinetic energy cutoff of 35 Ry. The 
ground state charge density was determined on a BZ grid of 14$\times$14$\times$14 
$\bm{k}$-points. Quasiparticle energies within the GW approximation 
for the self-energy operator\cite{HedinLundqvist70,*HybertsenLouie86,*CFM} 
were determined on a 6$\times$6$\times$6 grid and interpolated throughout the 
BZ through the use of the maximally-localized Wannier function 
formalism\cite{MarzariVanderbilt97,*SouzaMarzariVanderbilt01,*wannier90}.
We included 34 electronic bands in the coarse-grid calculation and 
extracted 26 Wannier functions, which reproduce the LDA bandstructure 
10 eV below and 30 eV above the Fermi level. The interpolated quasiparticle 
band structure of silicon is shown in Figure~\ref{fig:fig1}. The indirect 
(1.3 eV) and direct (3.3 eV) quasiparticle band gaps are in good agreement 
with previous calculations\cite{HybertsenLouie86} and experiment.
The same formalism has been used to interpolate the velocity matrix 
elements\cite{PhysRevB.62.4927,PhysRevB.75.195121,*PhysRevB.74.195118},
including the 
renormalization required
\cite{PhysRevLett.63.1719} after the GW corrections.
The
real ($\varepsilon_1$) and imaginary ($\varepsilon_2$)
parts of the dielectric function and the refractive index
due to direct transitions,
required in Eq.~\ref{eq:alpha2} 
to determine the absorption coefficient,
were also determined at the quasiparticle level for a range of photon 
frequencies.
Lattice dynamics are calculated 
using density functional perturbation theory\cite{Baroni_et_al_2001}. The 
electron-phonon coupling matrix elements are calculated on the same coarse 
grid of electronic points, while the dynamical matrices and phonon-potential 
perturbations are calculated on a 6$\times$6$\times$6 grid of momentum-space 
vectors\cite{PhysRevB.76.165108,*PhysRevLett.99.086804}
and interpolated for arbitrary pairs of points
in the first BZ
using the \textsc{epw} 
code\cite{Noffsinger20102140}.
For the calculations of the velocity and electron-phonon coupling matrix elements
we used the LDA wave functions, because their overlap with the GW-corrected ones is better than 99.9\%\cite{HybertsenLouie86}.

\begin{figure*}
\includegraphics[scale=1.15]{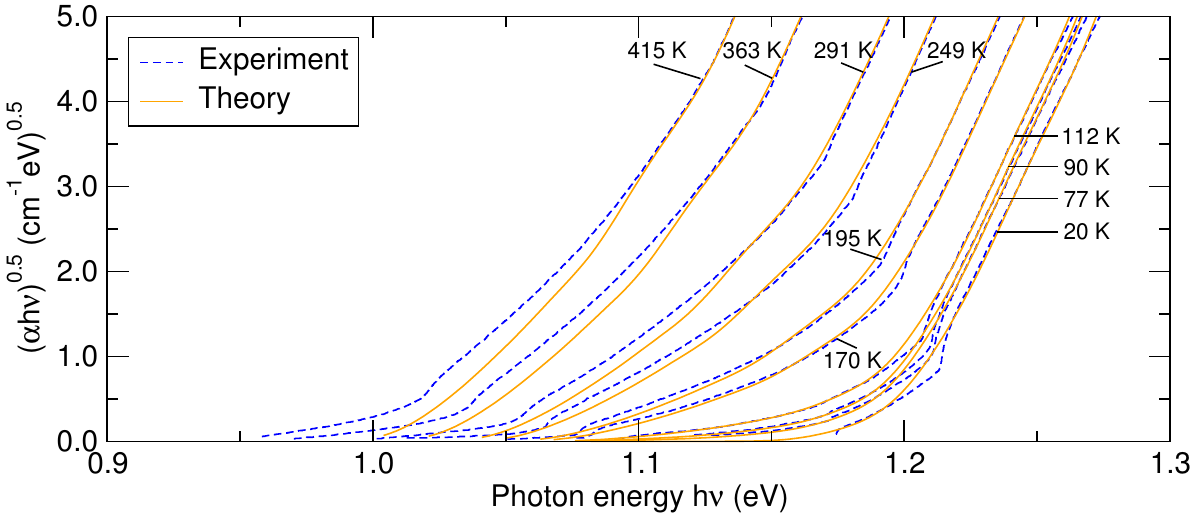}
\caption{\label{fig:fig2} Onset of the phonon-assisted optical 
absorption in silicon, as a function of photon energy and temperature. 
The theoretical results (solid lines) are in good agreement
with experiment (dashed lines). Experimental data are from Ref.~\onlinecite{PhysRev.111.1245}.
The theoretical curves have been shifted horizontally to match the onset of the experimental spectra.
}
\end{figure*}

Phonon-assisted optical absorption in indirect-band-gap semiconductors 
occurs for photons with energies greater than the indirect band gap minus 
(plus) the energy of the phonon absorbed (emitted) to 
assist the transition. The onset of indirect absorption is
calculated over a wide range of temperatures in bulk silicon through 
Eq.~\ref{eq:alpha2} and the results are shown in Fig.~\ref{fig:fig2}.
%
Each curve displays a characteristic knee, arising 
from the different energy onsets of the phonon-absorption and 
phonon-emission terms, which becomes smoother with increasing temperature.
The calculated data are in good agreement with experimental 
results\cite{PhysRev.111.1245} for all temperatures measured.
For these calculations, we used fine grids of 40$\times$40$\times$40 
for the $\bm{k}$ and $\bm{q}$ sums in 
Eq.~\ref{eq:alpha2}, respectively. These fine grids yield converged 
optical spectra with an energy resolution of 14 meV,
which is quite small and necessary to resolve the fine features near 
the absorption onset.
Although the experimental data near the edge can be fit with 
simple parameterized forms\cite{PhysRev.111.1245}, to our knowledge 
they have not been calculated entirely from first principles previously.

\begin{figure}
\includegraphics[scale=0.3]{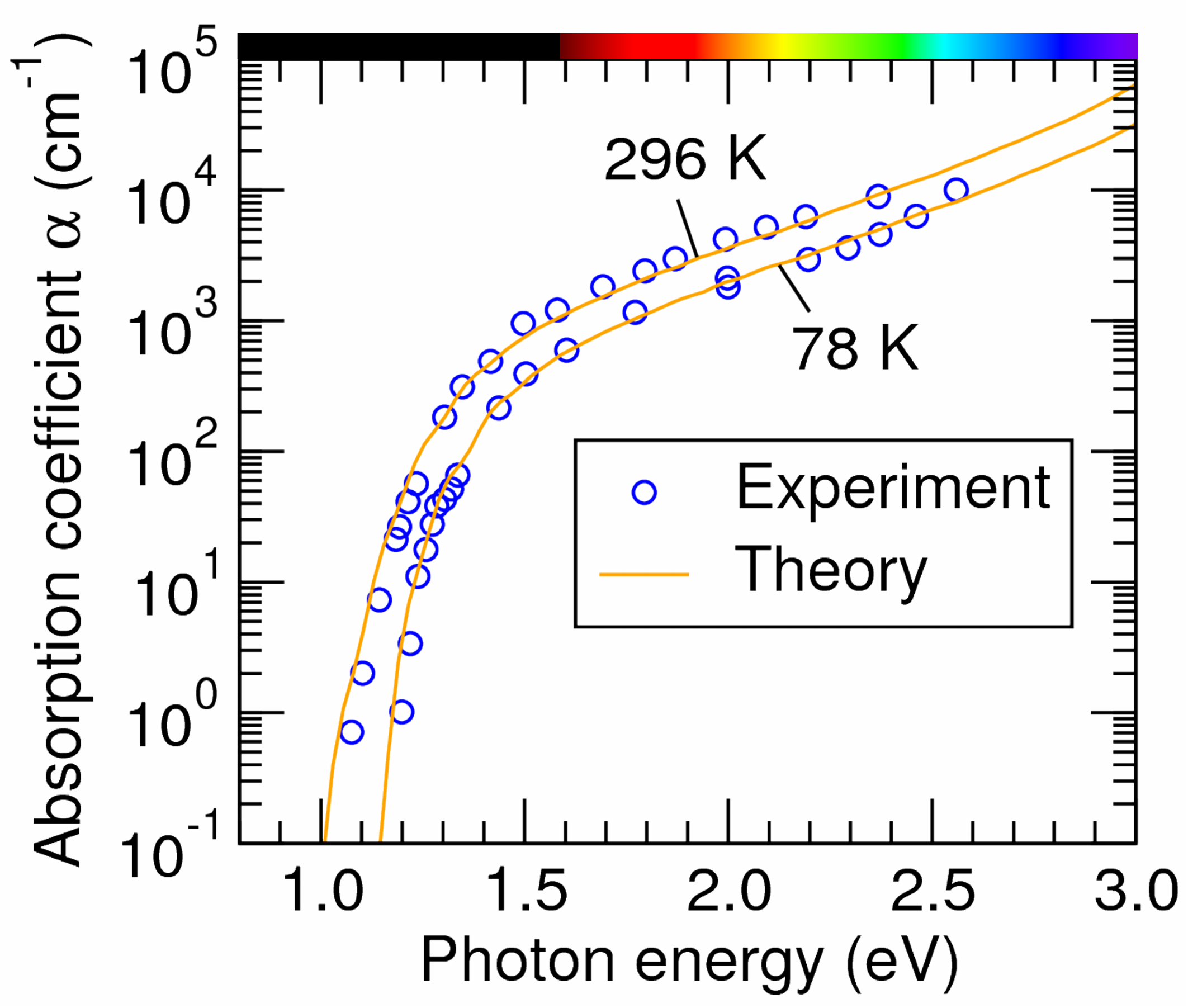}
\caption{ \label{fig:fig3} Calculated (solid lines) and experimental (circles) 
absorption coefficient of silicon in the energy range between the indirect and 
direct gaps, for two temperatures. Experimental data are from Ref.~\onlinecite{PhysRev.109.695}.
The theoretical spectra have been shifted to match the experimental absorption onset.
}
\end{figure}

In addition to the absorption onset, we are interested in the phonon-assisted 
absorption spectrum in the energy range between the indirect and direct band 
gaps, covering the visible range. This spectral region involves transitions 
between valence and conduction band states away from the band extrema and, as 
a consequence, cannot be modeled with simple parameterized forms. On the 
contrary, because of the large number of electronic states and phonon modes involved,
first-principles calculations are the only computational tool that can access 
this spectral region. The interpolation of the \textit{ab initio} quantities 
within the Wannier-Fourier formalism makes the calculation 
feasible on modern computers. The calculated spectra with an energy resolution 
of 30 meV (Fig.~\ref{fig:fig3}) converge with zone-sums of 24$\times$24$\times$24 
electronic and 24$\times$24$\times$24 phonon points.
The imaginary part of the 
electron self-energy for the intermediate electronic states was set to a 
constant value (100 meV). However, the calculated data are not very sensitive 
to the particular value of the quasiparticle lifetime for photon energies in 
this spectral region.

To facilitate comparison with experiment,
the theoretical absorption spectra of Figs.~\ref{fig:fig2} and \ref{fig:fig3} 
have been rigidly shifted to the left along the energy axis by 
0.15--0.23 eV in order to match the onset of the experimental absorption curves.
This shift is needed to account for 
the numerical difference between the calculated and experimental band gap,
and for finite-temperature effects on the quasiparticle energies
which we have not considered explicitly.
Although the GW method is the most accurate first-principles computational formalism
for the calculation of quasiparticle properties presently available,
it typically yields absolute quasiparticle energies accurate to 0.1 eV.
In our particular case, we found that the calculated band gap is also within this error bar 
larger than the experimentally measured value.
We note that no other first-principles 
method is presently available to guarantee a more accurate result.
Moreover, the only temperature dependence we considered in our calculations is for
the phonon occupation numbers.
However, the quasiparticle energies themselves are temperature-dependent
because of additional finite-temperature
effects, such as lattice expansion and electron-phonon renormalization
\cite{PhysRevB.76.165108,*PhysRevLett.99.086804,Claudia2008,*PhysRevB.79.245103,Bohnen2008,PhysRevB.31.2163,Feliciano_el_phon_gap_correction,*ANDP:ANDP201000100,Marini2008}.
We found that the thermal-expansion correction to the indirect band gap is small 
and amounts only to a 2.5 meV increase of the LDA band gap as the lattice constant
increases from the 0 K to the 400 K value \cite{Reeber1996259}, in agreement with Ref. \onlinecite{PhysRevB.31.2163}.
On the other hand,
empirical pseudopotential calculations have shown that
electron-phonon renormalization effects are stronger
and decrease the band gap of silicon by approximately 0.05--0.1 eV
for temperatures in the range 0--400 K
\cite{PhysRevB.31.2163}.
We note that the determination of electron-phonon-coupling corrections to quasiparticle energies from first principles
is still a subject of ongoing research
\cite{Feliciano_el_phon_gap_correction,*ANDP:ANDP201000100}.
The cumulative effects of this electron-phonon band-gap renormalization (0.05--0.1 eV)
and the intrinsic accuracy of the GW method (order of 0.1 eV)
explain the difference between the onsets of the theoretical and experimental data that we need to 
take into account when comparing our calculated spectra to experiment.
Moreover, this electron-phonon coupling correction to the quasiparticle energies
may have an effect on the shape of the absorption spectra near the onset of indirect transitions in Fig.~\ref{fig:fig2}.

Although excitonic effects, arising from the electron-hole interaction, are potentially 
important for optical processes and in general affect the direct absorption spectra 
even for photon energies far from the absorption edge\cite{PhysRevB.62.4927}, 
we found that they are not as important for the case of indirect absorption.
Sharp features that appear near the onset of indirect absorption 
are attributed to excitonic effects\cite{PhysRev.111.1245,PhysRevLett.24.942}.
Our calculations, however, are based on quasiparticle theory and do not account for the electron-hole interaction
that gives rise to these excitonic features.
Nevertheless, the
calculated absorption spectra are in very good quantitative 
agreement with 
experimental data, pointing to a weaker role of the electron-hole Coulomb 
interaction for the case of indirect optical transitions.
This is probably because
the band-extrema wavefunctions in indirect-gap materials are 
located at different points of the BZ
and hence the wavefunction overlap, which determines the magnitude of 
the Coulomb interaction between them, is small. Therefore, the phonon-assisted 
spectra can to a large extend be explained at the quasiparticle level of 
theory, without the need to account for excitonic effects.

The computational 
formalism we developed is
based on first-principles methods
and can be used to 
study the fundamental 
physics of phonon-assisted absorption in materials in general.
It can complement experimental studies to shed light on the microscopic phonon-assisted transition mechanisms
and address questions that are not accessible by experimental techniques.
Moreover, the method can be used to analyze the phonon-mediated optical properties of 
technologically important materials for optoelectronic applications.
E.g., it can investigate the role of phonon-assisted optical processes in silicon photonics, or it can predict
the photovoltaic performance of indirect-band-gap materials.

In conclusion, we used a Wannier-Fourier interpolation technique to calculate 
the phonon-assisted optical absorption spectra of silicon at the quasiparticle 
level. The calculated spectra are in very good agreement with experimental 
measurements, both near the absorption onset and in the spectral region between 
the indirect and direct band gaps for any lattice temperature. The first-principles computational 
formalism is very general and can be used to study the fundamental physics of phonon-assisted
absorption, as well as the phonon-mediated optical properties of optoelectronic materials.

\begin{acknowledgments}
We thank F. Giustino, P. Zhang, G. Samsonidze, B. Malone, and C. Carbogno for useful discussions.
E. K. was supported as part of the Center for Energy Efficient Materials,
an Energy Frontier Research Center funded by the U.S. DOE, BES under Award Number DE-SC0001009.
Additional support was provided by the UCSB Solid State Lighting and Energy Center.
J. N. was supported by National Science Foundation Grant No. DMR10-1006184 and 
by the Director, Office of Science, Office of Basic Energy Sciences, Materials Sciences 
and Engineering Division, U.S. Department of Energy under Contract No. DE- AC02-05CH11231.  
The GW code and work are supported by NSF and the electron-phonon code and computations
are supported by the DOE.
Computational resources were provided by the CNSI Computing Facility under NSF grant No. CHE-0321368,
the DOE NERSC facility, and Teragrid.
\end{acknowledgments}

\end{document}